# Deep Learning based Security-Constrained Unit Commitment Considering Locational Frequency Stability in Low-Inertia Power Systems

Mingjian Tuo
*Student Member, IEEE*
Department of Electrical and Computer Engineering
University of Houston
Houston, TX, USA
mtuo@uh.edu

Xingpeng Li
*Senior Member, IEEE*
Department of Electrical and Computer Engineering
University of Houston
Houston, TX, USA
xli82@uh.edu

*Abstract*—With the goal of electricity system decarbonization, conventional synchronous generators are gradually replaced by converter-interfaced renewable generations. Such transition is causing concerns over system frequency and rate-of-change-of-frequency (RoCoF) security due to significant reduction in system inertia. Existing efforts are mostly derived from uniform system frequency response model which may fail to capture all characteristics of the systems. To ensure the locational frequency security, this paper presents a deep neural network (DNN) based RoCoF-constrained unit commitment (DNN-RCUC) model. RoCoF predictor is trained to predict the highest locational RoCoF based on a high-fidelity simulation dataset. Training samples are generated from models over various scenarios, which can avoid simulation divergence and system instability. The trained network is then reformulated into a set of mixed-integer linear constraints representing the locational RoCoF-limiting constraints in unit commitment. The proposed DNN-RCUC model is studied on the IEEE 24-bus system. Time domain simulation results on PSS/E demonstrate the effectiveness of the proposed algorithm.

## I. INTRODUCTION

Zero-carbon wind and solar energy sources dominate the new electricity generation installed over the past decades. Besides the increasing penetration of converter-based renewable energy sources (RES), the development of energy storage systems and high-voltage dc (HVDC) transmission systems have resulted in large-scale application of power-electronic devices [1]. As a result of this transition, current power systems may ultimately shift towards power systems with 100% renewable-based generation.

Traditionally, the characteristics of power systems is primarily dominated by synchronous generators (SGs). Due to the strong coupling between the synchronous generator's rotor and the power system, the inertia stored in synchronous generator rotors plays an important role in regulating the power system frequency dynamics. With more generation coming from converter-based resources, insufficient inertia would be a main challenge for power systems stability [2]. Moreover, due to the retirement and replacement of conventional generation, the system kinetic energy is decreasing significantly, leaving the system more likely to subject to high rate of change of frequency (RoCoF) and large frequency excursion. When RoCoF violates the necessary industrial control and operation standards, protection devices would disconnect generators from the grid [3]. In particular, insufficient inertia exacerbates the need for fast frequency response services to secure frequency stability [4]. Altering RoCoF protection and enabling emulated inertia measures were found to be the most effective in reducing the frequency stability risk of future converter-based power system in Ireland [5].

As a mitigation strategy in the category of preventive actions, several transmission system operators (TSO) impose extra RoCoF related constraints in the conventional unit commitment model to keep the minimum amount of synchronous inertia online [6]. Requirements for adequate frequency control of the electric power system were suggested by Federal Energy Regulatory Commission [7]. EirGrid has also introduced a synchronous inertial response (SIR) constraint to ensure that the available inertia does not fall below a static limit of 23,000 MWs in Ireland [8]. The Swedish TSO ordered the Oskarshamn Kraftgrupp to reduce its nuclear power output by 100 MW to mitigate the risk of loss of the power plant in Sweden [9].

Traditionally, by looking at the collective performance of all generators using a system equivalent model, frequency related constraints can be derived and incorporated into the optimization formulation. The system equivalent model-based frequency constrained stochastic economic dispatch is considered in [10]. Reference [11] incorporated frequency-related constraints into traditional unit commitment enforcing limitations on RoCoF that is derived from a uniform frequency response model. A mixed analytical-numerical approach based on multi-regions has been studied in [12], which investigated the model combining evolution of the center of inertia and certain inter-area oscillations. Despite the previous efforts of modelling system frequency response (SFR) in mathematical programming-based scheduling, they ignored the nodal inertial response and the impact of disturbance propagation [13].

Reference [14] evaluates a fast transient frequency stability assessment using a data-driven tool, based on deep neural network (DNN). The actual input-output feature data is utilized to train the network parameters, which effectively extracts system characteristics. A DNN-based trajectory constraint encoding framework is proposed in [15], which incorporates frequency nadir and stability characteristics related constraints against the worst-case contingency. However, none of these methods considers locational RoCoF security and the impact of generator aggregations on system stability is not handled well.

This paper is to address the aforementioned issues. The contributions of this paper are as follows. First, we proposed a novel DNN-based RoCoF-constrained unit commitment (DNN-RCUC) model to secure the system locational frequency stability against worst contingencies. The proposed DNN-RCUC model shows better performance in handling the conservativity issues. Secondly, a model-based data generation approach is introduced to generate practical and ideal cases for RoCoF predictor training. The proposed method covers vast ranges of operating conditions, being able to avoid divergent time domain simulations. In addition, infeasible cases that may degrade the RoCoF predictor's accuracy are filtered out. Thirdly, the impact of generator aggregations is considered in our case study, showing the proposed DNN-RCUC model can secure locational frequency stability in various conditions.

The remainder of this paper is organized as follows. Section II discusses the power system mathematical based model and DNN-based model thoroughly. Section III presents the methodology of model-based data generation and DNN-RCUC formulation. Section IV shows the simulation results. Section V presents the concluding remarks and future work.

## II. SYSTEM FREQUENCY DYNAMICS

### A. System Equivalent Model

The frequency of the power system is one of the most important metrics that indicate the system stability. Traditionally, the frequency is treated as unique of the whole system, which is derived from the system equivalent model extended from one-machine swing equation. The rotating inertia of a synchronous generator is equal to the stored energy $E_i$ in the rotors of the machine at nominal speed, which is defined as:

$$E_i = \frac{1}{2} J_i \omega_i^2 \quad (1)$$

The rotational inertia of a single shaft is commonly defined using its inertia constant and the rated apparent power [16]. For a single machine, the inertia constant is expressed as follows:

$$H_i = \frac{J_i \omega_i^2}{2 S_{B_i}} \quad (2)$$

where $H_i$ is the inertia constant of the generator in seconds; $J_i$ is the moment of inertia of the shaft in kg·m$^2$s; $S_{B_i}$ is the base power in MVA; and $\omega_i$ is the nominal rotational speed instead of the actual speed of the machine.

Power system inertia is defined as the total amount of rotational energy stored in all rotating synchronous units; dynamics of these generators' rotors are directly coupled with the grid electrical dynamics. It can be expressed as follows.

$$E_{sys} = \sum_{i=1}^{N} \frac{1}{2} J_i \omega_i^2 = \sum_{i=1}^{N} H_i S_{B_i} \quad (3)$$

For a single generator $i$, the swing equation is expressed as:

$$\frac{d\omega_i}{dt} = \frac{P_m - P_{load}}{2 H_i S_{B_i}} \omega_n \quad (4)$$

where $P_m$ is the mechanical power and $P_{load}$ is the load from the power system, while $\omega_n$ is the rated steady state frequency of the system. $d\omega_i/dt$ is more commonly known as RoCoF. The swing equation of the system equivalent model can be then applied to the whole grid [17]. After a disturbance of power mismatch occurrence, the system RoCoF related to the total system inertia can be defined as,

$$RE^{t_0} = \frac{-\Delta P}{2 H_{sys} S_B} \omega_n \quad (5)$$

where $\Delta P$ is the sudden change in active power in MW at $t=t_0$.

### B. Dynamic Model

Following a sudden change in load or a generation contingency, the dynamic behavior of the system frequency can be described using the swing equation of system equivalent single-machine representation,

$$P_m - P_e = M \frac{\partial \Delta \omega}{\partial t} + D \Delta \omega \quad (6)$$

where $M$ and $D$ are the aggregated system inertia constant and damping coefficient corresponding to the committed synchronous generators respectively. $P_m$ is the mechanical input power. $P_e$ is the electrical output power. However, only considering the dynamics of the equivalent model in systems would underestimate the actual need for frequency ancillary services, leading to higher locational RoCoF and larger regional frequency deviation than expected.

Using the topological information and the system parameters, the transmission network can be modeled as a graph consisting of nodes (buses) and edges (branches). The oscillatory behavior of each individual bus can be expressed as follows,

$$m_i \ddot{\theta}_i + d_i \dot{\theta}_i = P_{in,i} - \sum_{j=1}^{n} b_{ij} \sin(\theta_i - \theta_j), \quad (7)$$
$$i \in \{1, \dots, n\}$$

where $m_i$ and $d_i$ denote the inertia coefficients and damping ratio for node $i$ respectively, while $P_{in,i}$ denotes the power input. With inertia on certain nodes $m_i > 0$, it is an approximation model for the swing dynamics of high-voltage transmission network within a short period following the event [18]. A network-reduced model with $N$ generator buses can be obtained by eliminating passive load buses via Kron reduction [19]. The phase angle $\theta$ of generator buses can be expressed by the following dynamic equation,

$$M \ddot{\theta} + D \dot{\theta} = P - L \theta \quad (8)$$

where $M = \text{diag}(\{m_i\})$, $D = \text{diag}(\{d_i\})$. And for the Laplacian matrix $L$, its off-diagonal elements are $l_{ij} = -b_{ij} V_i^{(0)} V_j^{(0)}$, and diagonals are $l_{ij} = \sum_{j=1, j \neq i}^{n} b_{ij} V_i^{(0)} V_j^{(0)}$. Under the assumption of homogeneous inertia, the frequency deviations at bus $i$ can then be derived [18],

$$R_i(t) =$$

$$\frac{\Delta P e^{-\frac{\gamma t}{2}}}{2\pi m} \sum_{\alpha=1}^{N_g} \frac{\beta_{\alpha i}\beta_{\alpha b}}{\sqrt{\frac{\lambda_\alpha}{m} - \frac{\gamma^2}{4}}\Delta t} \left[ \begin{array}{c} e^{-\frac{\gamma \Delta t}{2}} \sin\left(\sqrt{\frac{\lambda_\alpha}{m} - \frac{\gamma^2}{4}}(t+\Delta t)\right) \\ -\sin\left(\sqrt{\frac{\lambda_\alpha}{m} - \frac{\gamma^2}{4}}t\right) \end{array} \right] \quad (9)$$

where $m$ denotes average inertia distribution on generator buses; and bus $b$ is where disturbance occurs. The ratio $\gamma$ of damping coefficient $d_i$ to inertia coefficient $m_i$ is considered constant.

*C. DNN Based RoCoF Prediction*

Prediction of the highest nodal RoCoF is quite challenging due to the non-linear nature of the power system. Although efforts like [10] improve the frequency dynamics model by including the state-of-the-art converter control schemes, model-based approaches are unable to capture the entire characteristics and incorporate high-order models. DNN has shown the ability to amend the limitations of model-based approaches [15].

Given the load forecast $d$ and RES forecast $r$, following a disturbance $\varpi_\tau$ at period $\tau$, the highest locational RoCoF value of the system $R_h$ is a function with respect to the disturbance level, disturbance location, system states, dispatch command, load condition and RES profile.

$$R_h = h^r(s_\tau, u_\tau, d_\tau, r_\tau, \varpi_\tau) \quad (10)$$

where $s_\tau$ denotes the system states, and $u_\tau$ is the generation dispatch at period $\tau$. Compared to the case of a sudden load increase, the loss of generation not only causes mismatch in system power balance but also degrades the system synchronous inertia, resulting in even higher frequency deviation and larger initial RoCoF. Based on our discussion regarding the impact of Fiedler mode on inertial response, disturbance $\varpi_\tau$ is defined to be the $G$-1 contingency, the magnitude of event is determined by the generator output power. The DNN based RoCoF predictor for $h^r$ is then expressed as,

$$\hat{R}_h = \hat{h}^r(x, W, b) \quad (11)$$

where $x$ is the feature vector, and $W$ and $b$ denote the parameters of a well-trained DNN. Within the transient time interval following a disturbance, the characteristics of all generators are assumed unchanged, thus, it is sufficient to select the status of each SG as input features. Both the magnitude and location of the disturbance will affect the inertial response. The generator status feature vector for sample $s$ is defined as follows,

$$u_s = [u_{1,s}, u_{2,s}, \cdots, u_{N_G,s}] \quad (12)$$

The disturbance feature vector is defined against the loss of largest generation, the magnitude is expressed as,

$$P_s^\varpi = \max_{g \in G}(P_{1,s}, \cdots, P_{2,s}, \cdots, P_{N_G,s}) \quad (13)$$

The location of the disturbance is then represented by the index of the generator,

$$g_s^\varpi = \arg\max_{g \in G}(P_{1,s}, \cdots, P_{2,s}, \cdots, P_{N_G,s}) \quad (14)$$

We encode the information of magnitude and location into the disturbance feature vector as,

$$\varpi_s^G = [0, \cdots, 0, \underbrace{P_s^\varpi}_{g_s^\varpi \text{th element}}, 0, \cdots, 0] \quad (15)$$

Laplacian matrix $L$ of the grid and Fiedler mode value depend on the power-angle characteristics, which are determined by the active power injection [13]. Thus, the active power injection of all SGs will be encoded into the feature vector.

$$P_s = [P_{1,s}, \cdots, P_{2,s}, \cdots, P_{N_G,s}] \quad (16)$$

The overall feature vector of a sample $s$ can be then defined as follows,

$$x_s = [u_s, \varpi_s^G, P_s] \quad (17)$$

Now consider a fully connected neural network with $N_L$ hidden layer. Each layer uses a rectified linear unit (ReLU) activation function as $\sigma(\cdot) = \max(\cdot, 0)$ and the output layer is a linear activation function. The predicted RoCoF can be expressed as follows,

$$z_1 = x_s W_1 + b_1 \quad (18a)$$

$$\hat{z}_m = z_{m-1} W_m + b_m \quad (18b)$$

$$z_m = \max(\hat{z}_m, 0) \quad (18c)$$

$$R_{h,s} = z_{N_L} W_{N_L+1} + b_{N_L+1} \quad (18b)$$

where $W_m$ and $b_m$ represent the weight and bias for the $m$-th hidden layer, and $W_{N_L+1}$ and $b_{N_L+1}$ represent the set of weight and bias of the output layer.

### III. METHODOLOGY

*A. Model-based Data Generation*

Wide-range space of all power injections is utilized in [15] to ensure reliability under vast ranges of operating conditions. However, such random injections may lead to divergence during the simulation initialization process. If time-domain simulation is initialized successfully, the transient stability may still be subject to system oscillation mode as well as large rotor angle differences. It has also shown that many randomly sampled power injections are not stable under the worst-case disturbance even with stability predictor applied. In addition, stability predictor in random generation approach may also increase the computational burden and compromise the efficiency of the algorithm.

Unlike randomly data generation considering wide-range space of dispatching, a model-based systematic data generation approach is proposed to generate reasonable and representative data that will be used to train RoCoF Predictors. Training samples are generated from models over various load and RES scenarios, traditional SCUC (T-SCUC) models and frequency constrained SCUC models are implemented in this process [20].

Given the load forecast and RES forecast, the T-SCUC is the base model generating dispatching samples. Objective function (19a) is to minimize the total system cost consisting of variable fuel costs, no-load costs, start-up costs, and reserve costs.

$$\min_\Phi \sum_{g \in G} \sum_{t \in T} (c_g P_{g,t} + c_g^{NL} u_{g,t} + c_g^{SU} v_{g,t} + c_g^{RE} r_{g,t}) \quad (19a)$$

$$\sum_{g \in G} P_{g,t} + \sum_{k \in K(n-)} P_{g,t} - \sum_{k \in K(n+)} P_{g,t} - D_{n,t} \quad (19b)$$
$$+ E_{n,t} = 0, \forall n, t$$

$$P_{k,t} - b_k(\theta_{n,t} - \theta_{m,t}) = 0, \forall k, t \quad (19c)$$

$$-P_k^{max} \leq P_{k,t} \leq P_k^{max}, \forall k, t \quad (19d)$$

$$P_g^{min} u_{g,t} \leq P_{g,t}, \forall g, t \quad (19e)$$

$$P_{g,t} + r_{g,t} \leq u_{g,t} P_g^{max}, \forall g, t \quad (19f)$$

$$0 \leq r_{g,t} \leq R_g^{re} u_{g,t}, \forall g, t \quad (19g)$$

$$\sum_{j \in G} r_{j,t} \geq P_{g,t} + r_{g,t}, \forall g, t \quad (19h)$$

$$P_{g,t} - P_{g,t-1} \leq R_g^{hr}, \forall g, t \quad (19i)$$

$$P_{g,t-1} - P_{g,t} \leq R_g^{hr}, \forall g, t \quad (19j)$$

$$v_{g,t} \geq u_{g,t} - u_{g,t-1}, \forall g, t \quad (19k)$$

$$v_{g,t+1} \leq 1 - u_{g,t} \quad \forall g, t \leq nT - 1 \quad (19l)$$

$$v_{g,t} \leq u_{g,t} \quad \forall g, t \quad (19m)$$

$$v_{g,t} \in \{0,1\}, \forall g, t \quad (19n)$$

$$u_{g,t} \in \{0,1\}, \forall g, t \quad (19o)$$

The T-SCUC model includes various constraints (19b)-(19o). Equation (19b) enforces the nodal power balance. Network power flows are calculated in (19c) and are restricted by the transmission capacity as shown in (19d). The scheduled energy production and generation reserves are bounded by unit generation capacity and ramping rate (19e)-(19j). As defined in (19h), the reserve requirements ensure the reserve is sufficient to cover any loss of a single generator. The start-up status and on/off status of conventional units are defined as binary variables (19k)-(19o).

As mentioned before, two RoCoF constrained security constrained unit commitment (SCUC) models are utilized. It is worth pointing out that the primary response is neglected in the formulation, without affecting the inertial response and the fundamental findings of this work. For system equivalent model based RoCoF constrained security constrained unit commitment (ERC-SCUC) model, constraint (19p) is introduced to guarantee generator frequency stability considering the relative location to the potential $G - 1$ contingency [11].

$$RE^{t_0}(\Delta P, H_{sys}, S_B) \leq -\text{RoCoF}_{\text{lim}}, \quad \forall g, t, \quad (19p)$$

The location based RoCoF constrained SCUC (LRC-SCUC) introduces locational RoCoF constraints based on the definition of local buses $N_{loc}$ and non-local buses $N_{n\text{-}loc}$. Constraints (19q) and (19r) ensure system stability by imposing limit on locational RoCoF over all buses under all $G - 1$ contingency.

$$R_n^{T_1}(\Delta P, H_{sys}, S_B) \leq -RoCoF_{lim}, \forall n \in N_{loc}, g, t \quad (19q)$$

$$R_n^{T_2}(\Delta P, H_{sys}, S_B) \leq -RoCoF_{lim}, \forall n \in N_{n\text{-}loc}, g, t \quad (19r)$$

The constraints on RoCoF for locational frequency dynamics are nonlinear. In order to incorporate these frequency-related constraints into the proposed LRC-SCUC model, a linear approximation method is introduced. The SCUC models used for data generation are summarized in TABLE I, the detail of all models is presented in [20].

TABLE I
DIFFERENT SCUC MODELS

| Model | Objective Function | Shared Constraints | Unique Constraints |
|---|---|---|---|
| T-SCUC | (19a) | (19b)-(19o) | None |
| ERC-SCUC | | | (19p) |
| LRC-SCUC | | | (19q)-(19r) |

B. DNN Linearization

To encode the DNN into the MILP SCUC problem, decision variables are introduced to build the disturbance feature vector. binary variable $\lambda_{g,t}^G$ is used to indicate the status of largest output power of generator $g$ in scheduling period $t$, a big-M method is introduced to express the disturbance vector,

$$P_{\rho,t} - P_{g,t} \leq M(1 - \lambda_{g,t}^G), \forall \rho, g, t \quad (20)$$

$$\sum_{g \in G} \lambda_{g,t}^G = 1 \ \forall t \quad (21)$$

where $M$ is a big positive number. Equation (20) enforces $\lambda_{g,t}^G$ to be zero is the dispatched output power of other generators $\rho$ is larger than generator $g$ at period $t$, while (21) limit the number of potential largest generator to be one at one period. When generator $g$ has the largest output power, (20) and (21) would set the status of largest output power $\lambda_{g,t}^G$ to be 1. To express the magnitude of disturbance, variable $\varepsilon_{g,t}$ is defined as the indicators of disturbance,

$$\varepsilon_{g,t} - P_{g,t} \geq -M(1 - \lambda_{g,t}^G), \forall g, t, \quad (22)$$

$$\varepsilon_{g,t} - P_{g,t} \leq M(1 - \lambda_{g,t}^G), \forall g, t, \quad (23)$$

$$0 \leq \varepsilon_{g,t} \leq M\lambda_{g,t}^G, \forall g, t, \quad (24)$$

Thus, the input feature vector can be expressed as follows,

$$x_t = [u_{1,t}, \cdots, u_{N_G,t}, \varepsilon_{g,t}, \cdots, \varepsilon_{N_G,t}, P_{1,s}, \cdots, P_{N_G,s}] \quad (25)$$

RoCoF-limiting constraints can be derived from the pre-trained RoCoF predictor $\hat{h}^r(x, W, b)$, the nonlinear constraints can be incorporated into MILP problems by introducing auxiliary binary variables $a$. The reformulation of DNN-based RoCoF predictor is as follows,

$$z_{1,t}^\alpha = x_t W_1^\alpha + b_1^\alpha, \forall t, \quad (26a)$$

$$\hat{z}_{m,t}^\alpha = z_{m-1,t}^\alpha W_m^\alpha + b_m^\alpha, \forall m, \forall t, \quad (26b)$$

$$z_{m[l],t} \leq \hat{z}_{m[l],t} - M(1 - a_{m[l],t}), \forall m, l, t \quad (26c)$$

$$z_{m[l],t} \geq \hat{z}_{m[l],t}^\alpha, \forall m, \forall l, \forall t, \quad (26d)$$

$$z_{m[l],t} \leq M a_{m[l],t}, \forall m, \forall l, \forall t, \quad (26e)$$

$$z_{m[l],t} \geq 0, \forall m, \forall l, \forall t, \quad (26f)$$

$$a_{m[l],t} \in \{0,1\}, \forall m, \forall l, \forall t, \quad (26g)$$

$$\hat{R}_{h,t} = z_{L,t} W_{L+1} + b_{L+1}, \forall t, \quad (26h)$$

Then the RoCoF related constrained can be formulated as,

$$\hat{R}_{h,t} \leq -\text{RoCoF}_{\text{lim}}, \forall t, \quad (26i)$$

## IV. CASE STUDIES

A case study on IEEE 24-bus system [21] is provided to demonstrate the effectiveness of the proposed methods. This test system contains 24 buses, 33 generators and 38 lines, which also considers decarbonized generation characterized by wind power. The base case has a total demand from 1,195 MW to a peak of 2,116 MW. To ensure the practicality of the dataset and the generality of the trained model, load profile and RES profile are sampled based on Gaussian distribution while the deviation of means value ranges from [-20%, 20%] of the based value. The mathematical model-based data generation is operated in Python using Pyomo [22]. Regarding post-contingency frequency limits, RoCoF must be higher than -0.5Hz/s to avoid the tripping of RoCoF-sensitive protection relays, and the optimality gap is set to 0.1%. The PSS/E software is used for time domain simulation and labeling process, and full-scale models with detailed generator dynamics are implemented for more realistic data. GENROU for the synchronous generators; IEEEX1 for the excitation system; IEESGO for the turbine governor; PSS2A for the power system stabilizer. The computer with Intel® Xeon(R) W-2195 CPU @ 2.30GHz and 128 GB of RAM was utilized to conduct the numerical simulations.

### A. Predictor Training

The base vector has a dimension of 99. For the DNN layers, the number of neurons is set 10 for each layer. Rectified linear unit (ReLU) is used as the activation function. The training was operated in batches of 32 data points. An MSE based dynamic learning rate strategy is used for the training. Learning rate schedule is applied in the training process by reducing the learning rate accordingly, the factor by which the learning rate will be reduced is set to 0.5 and the patience value is set to 10 epochs. A total of 9,600 samples were collected based on strategies proposed in previous section. The entire dataset is divided into two subsets: 7,680 samples (80%) for training and 1,920 samples (20%) for validation.

TABLE II shows the validation accuracy of the RoCoF predictor under different tolerances. The validation accuracy is 99.27% with 10% tolerance, implying high performance of the trained model. It should be noted that the accuracy is still above 93.55% even with a small tolerance of 5%, indicating the robustness of the trained predictor.

TABLE II
VALIDATION ACCURACY of the Proposed DNN-based ROCOF PREDICTOR

| Tolerance | 10% | 9% | 8% | 7% | 6% | 5% |
|---|---|---|---|---|---|---|
| Accuracy | 99.3% | 99.0% | 98.5% | 96.6% | 95.5% | 93.6% |

### B. DNN-RCUC Results

The forecast load and wind power for test case are plotted in Fig. 1 and Fig. 2. Due to the computational efficiency, one interval in the simulation process represents 4 hours in periods. The test case has a demand ranging from 1,633 MW to a peak of 1,853 MW. The peak wind generation is 266 MW. All three RoCoF-constrained models are tested on the same test case.

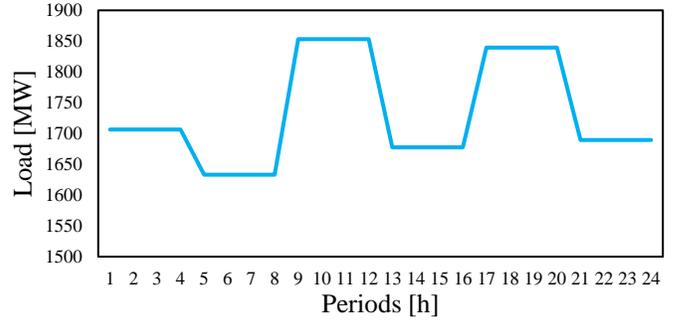

Fig. 1. Load profile of the IEEE 24-bus system.

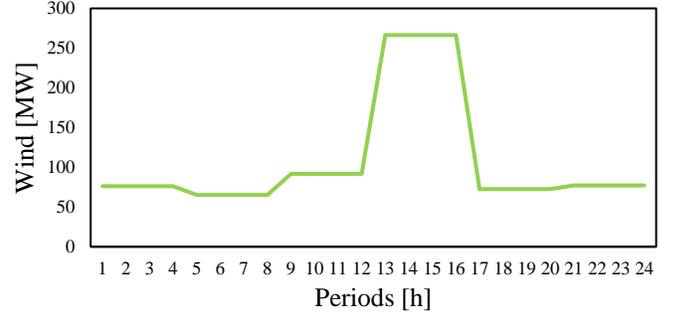

Fig. 2. Wind generation of the IEEE 24-bus system.

TABLE III compares the unit commitment results of the proposed DNN-based RCUC model and the two mathematical-based SCUC models. It can be observed that the proposed DNN-RCUC has the highest operational cost among all three models; the extra cost results from the efforts in handling the generator aggregation situations, which would be discussed later. On the other hand, the reserve cost is less in LRC-SCUC and DNN-RCUC models. For DNN-RCUC case, the total reserve cost is $46,608, which is slightly lower than the cost of LRC-SCUC model. Additional synchronous machines are committed to cover the loss of largest generation for DNN-RCUC model, which accordingly increases the operation cost as well as the start-up cost.

TABLE III
SCUC COSTS [$] UNDER DIFFERENT MODELS

| Model | Total | Start-up | Operation | Reserves |
|---|---|---|---|---|
| ERC-SCUC | 540,128 | 28,536 | 443,304 | 68,288 |
| LRC-SCUC | 732,864 | 40,372 | 645,884 | 46,608 |
| DNN-RCUC | 1,093,175 | 48,101 | 1,033,759 | 45,260 |

Additionally, we run the dynamic simulation of G-1 contingency for all three models when the system netload is the lowest. The loss of largest generation at this period is more likely to result in highest RoCoF and largest system deviations due to least synchronous generators online [11]. The highest RoCoF of three cases are listed in TABLE IV. Although with system equivalent model-based RoCoF constraints, ERC-SCUC model still cannot ensure system RoCoF security under such situation. The highest RoCoF of LRC-SCUC model is 0.3920 Hz/s, which gives a relatively high RoCoF violation gap at -21.60% below limit. The proposed DNN-RCUC has a highest RoCoF of 0.4952 Hz/s following the loss of largest generation. From TABLE III and Fig.3, it can be concluded that the proposed DNN-RCUC model can secure the system with minimal RoCoF violation gap while LRC-SCUC leads to conservative results.

TABLE IV
HIGHEST ROCOF OF DIFFERENT MODELS

| Model | ERC-SCUC | LRC-SCUC | DNN-RCUC |
|---|---|---|---|
| Highest RoCoF [Hz/s] | 0.6127 | 0.3920 | 0.4952 |

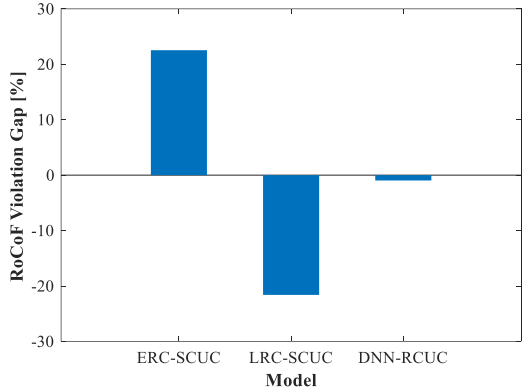

Fig. 3. RoCoF violation gaps for different cases.

Furthermore, the stability issue of generator aggregation is investigated [15]. In this work, we run time domain simulation of two locational RoCoF constrained models in the scenario considering the aggregation of generators on bus 23. Following the loss of the largest generation at hour 11, the highest RoCoF of two models are compared in TABLE IV. It can be observed that in the case of LRC-SCUC model, trip of the largest generation on bus 23 may cause generator on the same event bus violates the pre-specified RoCoF limit 0.5 Hz/s easily. When the constraints based on DNN are implemented, the system RoCoF can be secured just below the threshold.

TABLE IV
HIGHEST ROCOF OF DIFFERENT MODELS CONSIDERING GENRATOR AGGREGATIONS

| Model | LRC-SCUC | DNN-RCUC |
|---|---|---|
| Highest RoCoF [Hz/s] | 0.8125 | 0.4993 |

## V. CONCLUSIONS

In this paper, we comprehensively discussed the mathematical model and machine learning model for limiting RoCoF. Mathematical model based RCUC approaches may not capture all characteristics in various conditions. DNN-based RoCoF predictor is first constructed, and the constraints derived from the well-trained predictor are then incorporated into the RCUC model to ensure system stability. In addition, the proposed model-based data generation approach can avoid divergence in time domain simulation, which often occurs with random data generation method.

The simulation results on the IEEE 24-bus system indicate that the incorporation of DNN-based RoCoF constraints can improve power system inertial responses. Our proposed DNN-RCUC method has been proved to handle the condition of generator aggregations well. Future work on this topic should explore how to efficiently incorporate such complex DNN network into SCUC formulation.

REFERENCES

[1] L. Huang, H. Xin, Z. Wang, "Damping Low-Frequency Oscillations Through VSC-HVDC Stations Operated as Virtual Synchronous Machines," *IEEE Trans. Power Electron.*, vol. 34, no. 6, pp. 5803-5818, 2019.
[2] Mingjian Tuo, and Xingpeng Li, "Optimal Allocation of Virtual Inertia Devices for Enhancing Frequency Stability in Low-Inertia Power Systems", in *Proc. 53rd North Amer. Power Symp.*, Nov. 2021.
[3] F. Milano, F. Dörfler, G. Hug, D. J. Hill, and G. Verbič, "Foundations and challenges of low-inertia systems," in Power Systems Computation Conference (PSCC), Jun 2018.
[4] F. Teng, V. Trovato, and G. Strbac, "Stochastic Scheduling With Inertia Dependent Fast Frequency Response Requirements," IEEE Transactions on Power Systems, vol. 31, no. 2, pp. 1557–1566, 2016.
[5] J. O'Sullivan, A. Rogers, D. Flynn, P. Smith, and M. O'Malley," Studying the maximum instantaneous non-synchronous generation in an Island system—Frequency stability challenges in Ireland," IEEE Trans. Power Syst., vol. 29, no. 6, pp. 2943–2951, Nov. 2014.
[6] EirGrid and SONI, "DS3 System Services: Review TSO Recommendations," EirGrid, Dublin, Ireland, Tech. Rep., May 2013.
[7] J. H. Eto et al., "Use of frequency response metrics to assess the planning and operating requirements for reliable integration of variable renewable generation," Lawrence Berkeley National Laboratory, Berkeley, CA, USA, Tech. Rep. LBNL-4142E, Jan. 2011.
[8] EirGrid and SONI, "Operational Constraints Update," EirGrid, Tech. Rep., March 2019.
[9] H. Chavez, R. Baldick, and S. Sharma, "Governor rate-constrained OPF for primary frequency control adequacy," IEEE Trans. Power Syst.,vol. 29, no. 3, pp. 1473–1480, 2014.
[10] Matthieu Paturet, Uros Markovic, Stefanos Delikaraoglou, Evangelos Vrettos, Petros Aristidou and Gabriela Hug, "Stochastic Unit Commiment in Low-Inertia Grids," *IEEE Trans. Power Syst.*, vol. 35, no. 5, pp. 3448-3458, Sept. 2020.
[11] M. Paturet, Economic valuation and pricing of inertia in inverter-dominated power systems, Master thesis, Swiss Federal Institute of Technology (ETH) Zurich (2019).
[12] Luis Badesa, Fei teng, and Goran Strbac, "Conditions for Regional Frequency Stability in Power System Scheduling-Part I: Theory," *IEEE Trans. Power Syst.*, Early Access.
[13] L. Pagnier and P. Jacquod, "Inertia location and slow network modes determine disturbance propagation in large-scale power grids," PLoS ONE 14, e0213550 (2019).
[14] Wen Y, Zhao R, Huang M, Guo C. Data-driven transient frequency stability assessment: A deep learning method with combined estimation-correction framework. Energy Convers Econ. 2020;1(3):198–209.
[15] Y. Zhang, H. Cui, J. Liu, F. Qiu, T. Hong, R. Yao, and F. Li, "Encoding frequency constraints in preventive unit commitment using deep learning with region-of-interest active sampling," IEEE Transactions on Power Systems, vol. 37, no. 3, pp. 1942–1955, 2022.
[16] Y. Bian, H. Wyman-Pain, F. Li, R. Bhakar, S. Mishra, and N. P. Padhy, "Demand side contributions for system inertia in the GB power system," *IEEE Transactions on Power Systems*, vol. 33, pp. 3521-3530, Jul. 2011.
[17] Mingjian Tuo and Xingpeng Li, "Dynamic Estimation of Power System Inertia Distribution Using Synchrophasor Measurements", 2020 52nd North American Power Symposium (NAPS), Apr. 2021.
[18] M. Tyloo, L. Pagnier, and P. Jacquod, "The key player problem in complex oscillator networks and electric power grids: Resistance centralities identify local vulnerabilities," Sci. Adv., vol. 5, no. 11, 2019, Art. No eaaw8359.
[19] F. Dörfler and F. Bullo, "Kron reduction of graphs with applications to electrical networks," *IEEE Trans. Circuits Syst. I: Regular Papers* vol. 60, no. 1, pp. 150–163, Jan. 2013.
[20] Mingjian Tuo and Xingpeng Li, "Security-Constrained Unit Commitment Considering Locational Frequency Stability in Low-Inertia Power Grids", *IEEE Trans. Power Syst.*, under review.
[21] Arun Venkatesh Ramesh, Xingpeng Li, and Kory W Hedman, "An accelerated-decomposition approach for security-constrained unit commitment with corrective network reconfiguration," *IEEE Trans. Power Syst.*, vol. 37, no. 2, pp. 887-900, Mar. 2022.
[22] Hart, William E., Carl Laird, Jean-Paul Watson, David L. Woodruff, Gabriel A. Hackebeil, Bethany L. Nicholson, and John D. Siirola. Pyomo – Optimization Modeling in Python. Springer, 2017.